\documentclass[aps,onecolumn,amssymb]{revtex4}
\usepackage{epsfig,amssymb,amsmath}

\newcommand{\be}{\begin{equation}}
\newcommand{\ee}{\end{equation}}

\begin{document}
\title{Quantum $k$-core conduction on the Bethe lattice}
\author{L. Cao$^1$ and J. M. Schwarz$^1$}
\affiliation{$^1$Physics Department, Syracuse University, Syracuse, NY 13244}
\date{\today}
\begin{abstract}
Classical and quantum conduction on a bond-diluted Bethe lattice is considered.  The bond dilution is subject to the constraint that every occupied bond must have at least $k-1$ neighboring occupied bonds, i.e. $k$-core diluted. In the classical case, we find the onset of conduction for $k=2$ is continuous, while for $k=3$, the onset of conduction is discontinuous with the geometric random first-order phase transition driving the conduction transition. In the quantum case, treating each occupied bond as a random scatterer, we find for $k=3$ that the random first-order phase
transition in the geometry also drives the onset of quantum
conduction giving rise to a new universality class of Anderson localization transitions.
\end{abstract}
\maketitle

\section{Introduction}
Theoretical study of the interplay between uncorrelated disorder and quantum mechanics began with the seminal work of P. W. Anderson more than fifty years ago~\cite{anderson}. Anderson predicted the existence of spatially localized single particle states provided there is sufficient disorder in the potential. This finding identified the possibility of a phase transition from conducting (extended states) to insulating (localized states) with increasing variance in the distribution of disorder.  The nature of the metal-insulator transition eventually became more transparent with the introduction of the phenomenological one-parameter scaling theory of localization~\cite{gangof4} based on ideas developed by Wegner~\cite{wegner1} and Thouless~\cite{thouless}. The scaling theory was subsequently fortified by a nonlinear sigma field theory and a perturbation theory near two-dimensions allowing for calculation of the set of exponents characterizing the continuous Anderson localization transition~\cite{wegner2,russians,vollhardt}.

There exists another approach to studying the interplay between quantum mechanics and disorder via the model of quantum site percolation~\cite{degennes}. Consider a binary alloy where the potential landscape is represented by two different energies randomly distributed throughout the system with probabilities $p$ and $1-p$ respectively.  In the limit where the energy difference approaches infinity, the quantum particle can only access (hop to) one of the two types of atoms. These randomly distributed accessible sites may or not span the system and, hence, affect the quantum conduction.  If the accessible sites do not span the system then surely the system is insulating, for example.

Some controversy surrounds the study of quantum percolation.  For instance, some argue that quantum percolation transition in the same universality class as the Anderson transition~\cite{travenec,kaneko}. However, others find evidence for a transition in two dimensions~\cite{meir,chang,islam}, contrary to the scaling theory. One issue that appears to be resolved to some extent in three dimensions is the fact that $p_q>p_c$, where $p_c$ signals the onset of the geometric percolation transition and $p_q$ signals the onset of extended single particle wavefunctions~\cite{germans}.

The Anderson model and quantum percolation have been analyzed on the Bethe lattice~\cite{abou,miller,harris1,harris2}.  The loopless structure of the Bethe lattice makes it amenable for analytic study, hence, it will be implemented here. For instance, Abou-Chacra and collaborators were able to obtain closed form expression for the breakdown of localized states in terms of the value of the potential, the width of the disorder and the coordination number of the lattice~\cite{abou}.  Harris~\cite{harris1,harris2} analyzed a bond version of quantum percolation and found that the exponent associated with the divergence of the average finite cluster size as the transition is approached from below is the same as classical percolation but with $p_q>p_c$.

Traditionally, the Anderson model and quantum percolation are models of quantum transport with short-ranged, uncorrelated disorder.  More recently, eigenfunction studies of power-law diluted chains have been conducted~\cite{lyra}. More specifically, the hopping probability of an occupied bond scales with, $r$, the distance along the chain, as $1/r^{1+\upsilon}$. As $\upsilon$ is decreased below $0.68$, extended states emerge presumably due to the fact that the system is becoming mean-field-like such that this result is not contrary to the scaling theory of localization.

\begin{figure}[bth]
\begin{center}
\includegraphics[width=8cm]{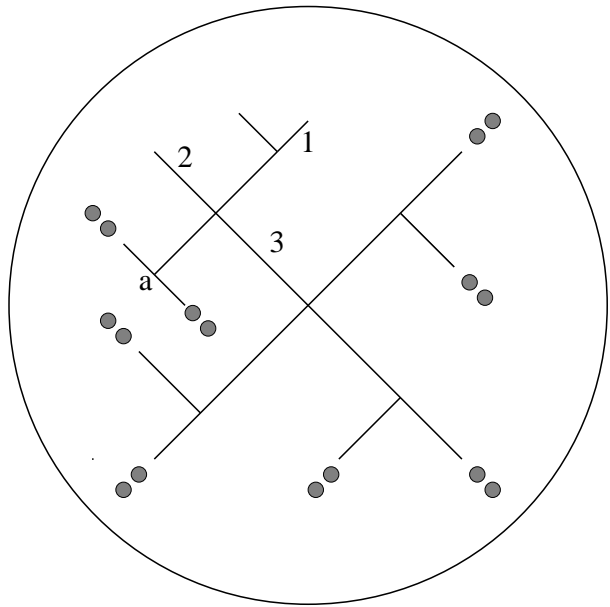}
\caption{Here $k=3$ and $z=4$.  The shaded circles denote branches that are $k-1$ connected to infinity.  The removal of bonds 1 and 2 eventually triggers removal of bond 3 and bonds emanating from vertex a, including the shaded circles. The remaining three branches emanating from the center site survive the removal process. }
\label{fig:bethe}
\end{center}
\end{figure}

How is the Anderson transition or the quantum percolation transition affected by other types of disorder---for example, correlated disorder where the correlations arise via local constraints on the occupation of bonds? The simplest model of correlated percolation is $k$-core bond percolation where every occupied bond must have at least $k-1$ occupied neighboring bonds~\cite{clr,wormald,slc,network,harris3}. To enforce this constraint, bonds are initially occupied independently and at random with probability $p$. Then, those occupied bonds with less than $k-1$ occupied neighboring bonds are rendered unoccupied.  This removal procedure proceeds recursively throughout the lattice until all occupied bonds satisfy the $k$-core constraint. Please see Fig.~\ref{fig:bethe} for an example on the Bethe lattice. 

As for the $k$-core percolation transition, in mean field, for $k\ge 3$, the fraction of occupied bonds in the spanning cluster, $P_\infty$, is finite at the transition. This result is to be contrasted with $k<3$, where the fraction of occupied bonds is zero at the transition. While the $k\ge 3$ transition is discontinuous in terms of the order parameter, $P_\infty$, there exists at least two diverging lengthscales exhibiting evidence of a random first-order phase transition~\cite{slc}. Therefore, the $k\ge 3$ represents a new universality class differing from the ordinary, uncorrelated percolation model.

Given this new universality class in the geometric percolation transition from disconnected to connected due to constraints on the disorder, let us return to the theme of the interplay between disorder and quantum mechanics. How does the random first-order phase transition in the geometry affect the onset of quantum conduction?  Could the discontinuous nature of the nature allow for $p_c=p_q$ as well as allow for the discontinuous onset of conduction, providing evidence for a new universality class in quantum localization transitions?

We will provide an answer to this question via analysis of quantum conduction on a $k$-core diluted Bethe lattice.  Before doing so, we present results of classical conduction on a $k$-core diluted Bethe lattice since we will implement some the machinery in the quantum limit as well. Note that the $k=1$ case has been analyzed previously by Stinchcombe~\cite{stinchcombe} and Kogut~\cite{kogut} has analyzed a site version of the $k=3$ case. Therefore, the paper is organized as follows: Section II provides the classical analysis, Section III provides the quantum analysis using Landauer conduction, and Section IV discusses the implications of our results.\\
\section{Classical $k$-core conduction}
\subsection{Geometry of $k$-core clusters}
Consider a seed vertex from which a lattice with coordination number $z$ and $N$ generations emerges. Each generation is constructed by recursively adding $z-1$ bonds to an $mth$ generation site forbidding the formation of loops to produce a rooted Bethe lattice.  Consider the missing $z$ bond of the seed site to survive the $k$-core removal process as specified in the introduction. Then the entire connected cluster of occupied bonds, each occupied with probability $p$, will survive the $k$-core removal procedure if each occupied bond in the $mth$ generation has $k-1$ occupied neighbors bonds in the $(m+1)th$ generation.
To determine the percolation properties of such geometrical structures, we define $R$, the probability that an arbitrarily chosen branch leaving a given site is not in an infinite $k$-core cluster with
\be
R=1-p+ p\sum_{n=0}^{k-2}{z-1 \choose n}R^{z-1-n}(1-R)^n.\label{eq:R}
\ee
The arbitrarily chosen branch is not part of an infinite $k$-core cluster if (a) the bond is not occupied or (b) the bond is occupied, but less than $k-1$ of its neighboring bonds are occupied.  For $k=2$, the equation for $R$ reduces to the ordinary bond percolation problem with $R$ decreasing continuously from unity just above the transition. In particular, for $z=3$ with $p-p_c=\epsilon<<1$,
\be
R = \left\{ \begin{array}{ll}
1 & \textrm{if ${\epsilon}<0$}\\
\frac{1}{p_c+{\epsilon}}-1=\frac{1}{1/2+{\epsilon}}-1\thicksim1-4{\epsilon}
& \textrm{if ${\epsilon}\geq0$}.
\end{array} \right.\label{eq:R1}
\ee For $k\ge 3$, however, $R$ jumps
discontinuously from unity at the transition such that $k\ge 3$
represents a different universality class from ordinary
percolation. For example, for $k=3,z=4$, 
\be 
R =
\left\{ \begin{array}{ll}
1 & \textrm{if ${\epsilon}<0$}\\
R_c-R_0{\epsilon}^{1/2} & \textrm{if ${\epsilon}\geq 0$}
\end{array} \right.
\label{eq:R2}
\ee where $p_c=\frac{8}{9}$,
$R_c=\frac{1}{4}$, and $R_0=\frac{9\sqrt{2}}{16}$.

\subsection{Classical conduction formulas}
Each occupied bond denotes a conductor with conductivity $\sigma_0>0$, while each unoccupied bond denotes a conductor with zero conductivity.  The probability of an arbitrarily chosen branch leaving a given site having a conductivity $\sigma$ is given by
\be
P(\sigma)=R\delta(\sigma)+(1-R)H(\sigma),
\label{eq:Psigma}\ee
where
\be
H({\sigma})=p\sum_{n=k-1}^{z-1}{z-1 \choose n}(1-R)^{n}R^{z-1-n}\int d{\sigma}_1...\, d{\sigma}_nH({\sigma}_1)...\,H({\sigma}_n){\delta}({\sigma}-S_n)
,\label{eq:hofsigma}\ee
with
\be
S_n=\frac{\sigma_0 T_n}{\sigma_0+T_n}
\label{eq:Sn}\ee
and
\be
T_n=\sum_{i=1}^{n}\sigma_i.
\label{eq:Tn}\ee
Note that $S_n$ assumes that the bond starting off the arbitrarily chosen branch is occupied. We have also invoked the self-similarity of the Bethe lattice such that $H(\sigma)$ is equivalent from one generation to the next. 

Once $H(\sigma)$ is determined for an arbitrarily chosen branch, the microscopic conductivity distribution for the system can be computed via
\be
\rho(\sigma)=\sum_{n=0}^{k-1}{z \choose n}(1-R)^{n}R^{z-n}\delta(\sigma)+\sum_{n=k}^{z}{z \choose n}(1-R)^nR^{z-n}\int d{\sigma}_1...\,d{\sigma}_nH({\sigma}_1)...\,H({\sigma}_n){\delta}({\sigma}-T_n).
\ee
The first term represents those realizations where the $k$-core criterion is not met for the $z$ branches and the second for those realizations otherwise.
We note that $P(\sigma)$ is normalized as is $H(\sigma)$ and $\rho(\sigma)$. Finally, the average microscopic conductivity of system, $<\sigma>$, is given by
\be
<\sigma>=\int d\sigma \sigma \rho(\sigma).
\ee

\subsection{$k=2,z=3$ case}
We first compute $H(\sigma)$ determined by
\be H({\sigma}) =
2pR\int d{\sigma}_1H({\sigma}_1){\delta}({\sigma}-S_1)+p(1-R)\int
d{\sigma}_1d{\sigma}_2H({\sigma}_1)H({\sigma}_2){\delta}({\sigma}-S_2)
\label{eq:hofsigmak2z3}
\ee
with $R$ given by Eq.~\ref{eq:R1}. 
One expects the onset of nonzero average conduction to occur at $p_c$, the occupation probability above which there exists a spanning cluster with probability unity.  Since the geometric transition is continuous, we expect that the conduction transition is also continuous. Therefore, we propose the ansatz, 
\be
H(\sigma)=\frac{1}{\epsilon^{\theta}}\bar{H}(\frac{\sigma}{\epsilon^\theta}). \ee
We set $\theta=1$. Inserting this ansatz into Eq.~\ref{eq:hofsigmak2z3} yields \be \bar{H}({\sigma}) =
2(\epsilon+\frac{1}{2})(1-4\epsilon)\int
d{\sigma}_1\bar{H}(\frac{{\sigma}_1}{\epsilon}){\delta}({\sigma}-\frac{\sigma_1\sigma_0}{\sigma_1+\sigma_0})+(\epsilon+\frac{1}{2})4\epsilon\int
d{\sigma}_1d{\sigma}_2\bar{H}(\frac{{\sigma}_1}{\epsilon})\bar{H}(\frac{{\sigma}_2}{\epsilon}){\delta}({\sigma}-\frac{(\sigma_1+\sigma_2)\sigma_0}{\sigma_1+\sigma_2+\sigma_0}).
\ee We let $\frac{\sigma}{\epsilon}=x$ and
$\frac{\sigma_i}{\epsilon}=x_i$ for $i=1,2$, multiply both
sides by $\exp(-sx)$, and integrate over $x$ from zero to infinity to arrive at \be
\mathcal{H}(s) = 2(\epsilon+\frac{1}{2})(1-4\epsilon)\int
dx_1\bar{H}(x_1)\exp(-\frac{s
x_1}{1+\frac{{\epsilon}x_1}{\sigma_0}})+(\epsilon+\frac{1}{2})4\epsilon\int
dx_1
dx_2\bar{H}(x_1)\bar{H}(x_2)\exp(-\frac{s(x_1+x_2)}{1+\frac{\epsilon}{\sigma_0}(x_1+x_2)}),
\ee where $\mathcal{H}(s)$ is the Laplace transform of
$\bar{H}(x)$.

\begin{figure}[bt]
\begin{center}
\includegraphics[width=8cm]{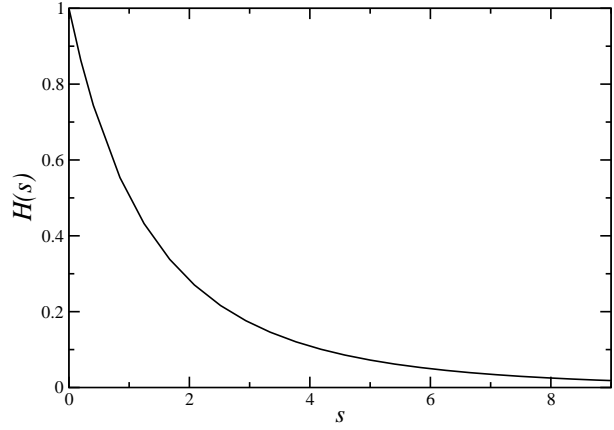}
\caption{$\mathcal{H}(s)$ for $\sigma_0=1/2$. }
\label{fig:diffeq}
\end{center}
\end{figure}

Expanding in powers of $\epsilon$ and collecting terms, the zeroth term in $\epsilon$ yields the identity
\be
\mathcal{H}(s)=\int dx_1 \bar{H}(x_1)\exp(-sx_1).
\ee
The first order term yields
\be
0=-2\int dx_1 \bar{H}(x_1)\exp(-sx_1)+\frac{s}{\sigma_0}\int dx_1 x_1^2 \bar{H}(x_1)\exp(-s x_1)+2\int dx_1 dx_2 \bar{H}(x_1)\bar{H}(x_2) \exp(-s(x_1+x_2))
\ee
leading to the differential equation,
\be
\mathcal{H}''(s)=2\frac{\sigma_0}{s}\mathcal{H}(s)(1-\mathcal{H}(s)).
\ee
As for the two boundary conditions, the normalization on $\bar{H}$ translates to $\mathcal{H}(0)=1$.  We also choose $\mathcal{H}(\infty)=0$.  We solve this system of equations numerically using Mathematica~\cite{mathematica}.  A result is shown in Figure~\ref{fig:diffeq}.   The large $s$ behavior is determined by $\mathcal{H}''(s)=2\frac{\sigma_0}{s}\mathcal{H}(s)$ with $\mathcal{H}(s)\sim \exp(-2\sqrt{2/\sigma_0}s^{1/2})$.  We note that the same form of the differential equation emerged from the analysis by Stinchcombe~\cite{stinchcombe} to second order in $\epsilon$. (See also similar analysis by Heinrichs and Kumar~\cite{kumar}.)

The above analysis justifies the ansatz for $H(\sigma)$ for $\epsilon<<1$. Inserting the ansatz into the expression for $<\sigma>$ to determine the scaling of $<\sigma>$ with $\epsilon$ yields
\be
<\sigma>=96\int dx x \bar{H}(x) \epsilon^3 +\mathcal{O}(\epsilon^4).
\ee
This result is to be contrasted with the result for $k=1$ analyzed by Stinchcombe~\cite{stinchcombe} where quadratic scaling with $\epsilon$ was computed. This difference is due to the fact that while the expression for $R$ is equivalent for $k=1$ and $k=2$, the fraction of bonds participating in the infinite cluster, $P_{\infty}$, scales differently with $\epsilon$ for $k=1$ and $k=2$.  For $k=1$, $P_{\infty}\sim \epsilon$, while for $k=2$, $P_{\infty}\sim \epsilon^2$ such that the conductivity should scale with $\epsilon$ as above. This scaling holds for larger $z$ as well. We note that the $k=2$ case is equivalent to biconnected percolation studied by Harris~\cite{harris4}.  Harris found that the exponent for $jth$-connectedness is given by $j\beta$, where $\beta=1$ is the order parameter exponent for ordinary percolation. Presumably, the conductivity exponent generalizes to $j\beta+1$ in the $jth$-connected case, which is different from $k$-core in that there is no culling process.

\subsection{$k=3, z=4$ case}
The equation for $H(\sigma)$ now reads
\be
H({\sigma})=3pR(1-R)\int
d{\sigma}_1d{\sigma}_2H({\sigma}_1)H({\sigma}_2){\delta}({\sigma}-S_2)+p(1-R)^2\int
d{\sigma}_1d{\sigma}_2d{\sigma}_3H({\sigma}_1)H({\sigma}_2)H({\sigma}_3){\delta}({\sigma}-S_3),\label{eq:hofsigmak3z4}
\ee
where $R$ is now given by Eq.~\ref{eq:R2}. Since there is a jump in $R$ at the transition for this value of $k$, one expects the transition in the conductivity to be discontinuous as well.  We propose the following scaling form for $H(\sigma)$:
\be
H({\sigma}) = \left\{ \begin{array}{ll}
0 & \textrm{if ${\epsilon}<0$}\\
H_c({\sigma})+{\epsilon}^{\lambda}K({\sigma}) & \textrm{if ${\epsilon}\geq0$.}
\end{array} \right.
\label{eq:ansatzk3z4}
\ee
The normalization on $H(\sigma)$ implies $\int d\sigma H_c(\sigma)=1$ and $\int d\sigma K(\sigma)=0$.

Using Eq.~\ref{eq:ansatzk3z4} and expanding in $\epsilon$, we arrive at
\be
H_c(\sigma)=\frac{1}{2}(\int d\sigma_1d\sigma_2 H_c(\sigma_1)H_c(\sigma_2)\delta(\sigma-S_2)+ \int d\sigma_1 d\sigma_2 d\sigma_3 H_c(\sigma_1)H_c(\sigma_2)H_c(\sigma_3)\delta(\sigma-S_3))
\label{eq:hcfull}
\ee
for the $\epsilon$ independent terms. If $\lambda>1/2$, then the terms of order $\epsilon^{1/2}$ yield
\be
 H_c(\sigma)=\int d\sigma_1d\sigma_2 H_c(\sigma_1)H_c(\sigma_2)\delta(\sigma-S_2)
\label{eq:hc1}
\ee
\and
\be
H_c(\sigma)=\int d\sigma_1 d\sigma_2 d\sigma_3 H_c(\sigma_1)H_c(\sigma_2)H_c(\sigma_3)\delta(\sigma-S_3)
\label{eq:hc2}
\ee
Equation~\ref{eq:hc1} yields solutions, $H_c(\sigma)=0,\delta(\sigma)$, or $\delta(\sigma-\frac{\sigma_0}{2})$.  The last solution is the only one allowed given the construction of $H_c(\sigma)$, however, this solution conflicts with the solution $H_c(\sigma)=\delta(\sigma-\frac{2}{3}\sigma_0)$ from Eq.~\ref{eq:hc2}. So, $\lambda\le \frac{1}{2}$.  If $\lambda<\frac{1}{2}$, the terms of order $\epsilon^\lambda$ imply
\be
K(\sigma)=\int d\sigma_1 d\sigma_2 H_c(\sigma_1) K(\sigma_2) \delta(\sigma-S_2)
+ \frac{3}{2}\int d\sigma_1 d\sigma_2 d\sigma_3 H_c(\sigma_1) H_c(\sigma_2) K(\sigma_3)\delta(\sigma-S_3).
\label{eq:kofsigma}
\ee
However, the linear integral equation for $K(\sigma)$ dictates that $K(\sigma)$ can be arbitarily rescaled by a factor. Since any physical solution should be unique, $K(\sigma)=0$ is the only solution. Therefore, we rule out $\lambda<\frac{1}{2}$.  This leaves $\lambda=\frac{1}{2}$. Then the terms of order $\epsilon^{1/2}$ are
\begin{eqnarray}
K(\sigma)=\int d\sigma_1 d\sigma_2 H_c(\sigma_1) [K(\sigma_2) -(3\sqrt{2}/4)H_c(\sigma_2)]\delta(\sigma-S_2)\nonumber\\
+ \int d\sigma_1 d\sigma_2 d\sigma_3 H_c(\sigma_1) H_c(\sigma_2) [(3\sqrt{2}/4)H_c(\sigma_3)+ (3/2)K(\sigma_3)]\delta(\sigma-S_3).
\end{eqnarray}

To justify the above ansatz, we must find nontrivial solutions for Eqs.~\ref{eq:hcfull} and ~\ref{eq:kofsigma} with $\lambda=\frac{1}{2}$.  We do so numerically by making an initial guess for $H_c(\sigma)$ and $K(\sigma)$ and solving the integral equations iteratively until both equations are satisfied within some tolerance.  This algorithm is first implemented for $H_c(\sigma)$ and then that numerical solution is used to solve for $K(\sigma)$. For the rest of the classical numerical analysis, we set $\sigma_0=1$. The initial guesses used are $H_c(\sigma)=\delta(\sigma-0.6)$ and $K(\sigma)=\delta(\sigma-0.6)-\delta(\sigma-0.3)$.  The domain is broken up into bins such that after integration, each bin being reassigned the maximum weight. 

\begin{figure}[bt]
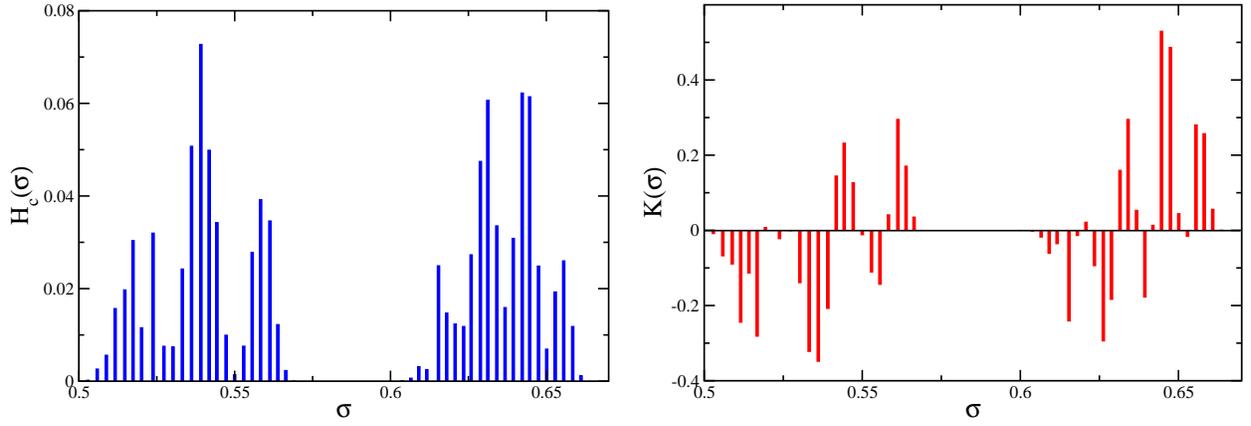

\begin{center}
\includegraphics[width=8cm]{k3.z4.classical.bond.hofsigma.eps}
\hspace{0.2cm}
\includegraphics[width=8cm]{k3.z4.classical.bond.kofsigma.eps}
\caption{Left: Plot of $H_c(\sigma)$ with binsize $0.0025$ for $k=3,z=4$ classical bond percolation.  Right: Plot of $K(\sigma)$ for the same conditions.
$H_c(\sigma)$ converges after 14 iterations and $K(\sigma)$ after 6 iterations.}
\label{fig:k3z4classicalbond}
\end{center}
\end{figure}

We plot the results for $H_c(\sigma)$ and $K(\sigma)$ with binsize $0.0025$ in Figure~\ref{fig:k3z4classicalbond}. Both solutions obey their respective constraints thereby justifying the initial ansatz. Using these numerical results we can also compute the average microscopic conductivity of the system near the transition.  More precisely,
\be
<\sigma>=c_1+c_2\epsilon^{1/2}
\ee
with $c_1=0.988$ and $c_2=1.945$. Note that the maximum conductivity of the system is $\frac{8}{3}$ in the fully occupied case.  \\

\underline{\it Monotone sequences}:
From Figure~\ref{fig:k3z4classicalbond}, the domain of $H_c(\sigma)$ appears to bounded away from zero and from the maximum value of $8/3$. As a check on our numerices, we determine these bounds with the following argument. Expanding $H_c({\sigma})$ as a sum of delta functions,
\be
H_c({\sigma})=\sum_{i}A_i{\delta}({\sigma}-a_i) \,\,\,\,\,\,\,\,\,\,0\leq A_i\leq 1, 0\leq a_i \leq \frac{8}{3},
\ee
and iterating Eq.~\ref{eq:hcfull} $n$ times yields the sequence, $a_{n,1}, a_{n,2},a_{n,3},...$, which is arranged from largest to smallest. The $(n+1)th$ iteration yields the sequence, $a_{n+1,1}, a_{n+1,2},a_{n+1,3},...$.

To determine the largest value of the $a_i$ after the $(n+1)th$ iteration, assume $a_{n,1}>a_{n,u}$ such that
\begin{eqnarray}
\frac{a_{n,1}+a_{n,u}}{1+a_{n,1}+a_{n,u}}\nonumber\\
&=&{}1-\frac{1}{a_{n,1}+a_{n,u}} \nonumber\\
&<&{}1-\frac{1}{a_{n,1}+a_{n,1}}\nonumber\\
&=&{}\frac{2a_{n,1}}{1+2a_{n,1}}.
\end{eqnarray}
We also have
\begin{eqnarray}
\frac{3a_{n,1}}{1+3a_{n,1}}\nonumber\\
&=&{}1-\frac{1}{3a_{n,1}+1}\nonumber\\
&>&{}1-\frac{1}{2a_{n,1}+1}\nonumber\\
&=&{}\frac{2a_{n,1}}{1+2a_{n,1}}.
\end{eqnarray}
So the largest value of $a_i$ after the $(n+1)th$
iteration is:
$\frac{3a_{n,1}}{1+3a_{n,1}}$.
Focusing on the largest values of each iteration, we form the sequence $Q$ with
\begin{eqnarray} \lefteqn {Q=\{{a_{1,1}, a_{2,1},a_{3,1},...\,a_{n,1},a_{n+1,1},...\}}}\nonumber\\
&=&{}\{a_{1,1},f(a_{1,1}),f(f(a_{1,1})),...,f^{(n-1)}(a_{1,1}),f^{(n)}(a_{1,1}),...\}\nonumber
\end{eqnarray}
Since
$a_{n,1}>f(a_{n,1})=a_{n+1,1}=\frac{3a_{n,1}}{1+a_{n,1}}$,
$a_{n,1}$, sequence $Q$ is a
monotonic decreasing sequence with a lower boundary of 0. Therefore, sequence $Q$ converges to a finite limit, $A$, determined by
\begin{displaymath}
A=f(A)=\frac{3A}{1+3A}
\end{displaymath}
with $A=2/3$. Therefore, sequence $Q$ converges to 2/3 with the largest value of $a_i=2/3$.

To obtain a lower bound on $a_i$, we construct another sequence,
\begin{eqnarray}
{Q'= \{a_{1,1}',g(a_{1,1}'),..., g^{(n-1)}(a_{1,1}'),g^{(n)}(a_{1,1}'),... \}},
\end{eqnarray}
where $a_{n,1}'$ is the smallest number after each iteration
and $g(x)=\frac{2x}{1+2x}$. Since
$a_{n,1}'<g(a_{n,1}')=\frac{2a_{n,1}'}{1+2a_{n,1}'}$, sequence $Q'$ is a monotonically increasing sequence with a boundary of unity so that the sequence approaches a finite limit, $A'$, with
\begin{displaymath}
A'=g(A')=\frac{2A'}{2A'+1}
\end{displaymath}
such that $A'=1/2$. Therefore, sequence $Q'$ converges to 1/2, i.e. the smallest value of $a_i$ is 1/2.

\begin{figure}[bth]
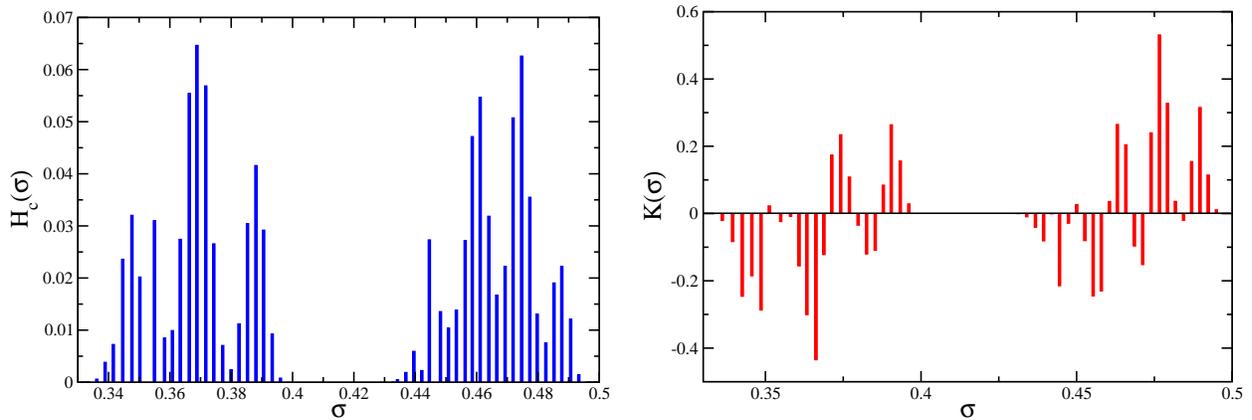

\begin{center}
\includegraphics[width=8cm]{k3.z4.classical.site.hofsigma.eps}
\hspace{0.2cm}
\includegraphics[width=8cm]{k3.z4.classical.site.kofsigma.eps}
\caption{Left: Plot of $H_c(\sigma)$ with binsize $0.0025$ for $k=3,z=4$ classical site percolation.  Right: Plot of $K(\sigma)$ for the same conditions.}
\label{fig:k3z4classicalsite}
\end{center}
\end{figure}

Consequently, the $a_i$s are confined between
1/2 and 2/3 for $H_c({\sigma})$ as demonstrated in the numerical analysis.  One can also extend this analysis to Kogut's $k=3$ site percolation analysis to demonstrate that the upper and lower bounds on $a_i$ are $1/2$ and $1/3$ respectively. (In the site formulation of the problem, occupied sites are surrounded by half-bonds with some conductivity and unoccupied sites by half-bonds of zero conductivity.) In Figure~\ref{fig:k3z4classicalsite} we plot $H_c(\sigma)$ and $K(\sigma)$ for the $k=3,z=4$ site percolation problem analyzed by Kogut to demonstrate the bounds. We note that the data suggests other gaps in the domains of these functions for the both the bond and site problem. Such gaps could indicate a fractal structure.

\vspace{1cm}

\section{Quantum $k$-core conduction}
\subsection{Quantum conduction formulae}
\begin{figure}[bt]
\begin{center}
\includegraphics[width=8cm]{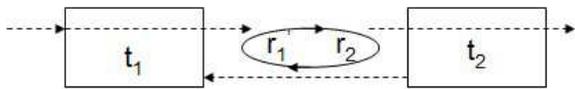}
\caption{Two quantum scatterers in series where $t_1$ and $r_1$ denote the transmission amplitude and reflection amplitude, respectively, for the first scatterer, and $t_2$ and $r_2$ the transmission and reflection amplitudes for the second. }
\label{fig:anderson}
\end{center}
\end{figure}

To examine quantum $k$-core conduction on a dilute Bethe lattice,
we require the conduction formulae for adding quantum resistors
in series and in parallel.  Anderson and
collaborators~\cite{anderson2} have derived the quantum equivalent of
Ohm's law for two quantum wires in series.
Their starting point is the Landauer approach to conductance, $g$, i.e. a scattering matrix approach~\cite{landauer}. Consider two scatterers in series. See 
Figure~\ref{fig:anderson}. The logarithm of the transmission probability of the two scatterer system is
\be
lnT=lnT_1+lnT_2-ln(1+R_1R_2-\sqrt{R_1R_2}cos(\theta)),\ee 
where $T_1=|t_1|^2$ with $t_1$ representing the transmission amplitude for the first scatterer, etc., and 
$\theta$ is the phase difference between the two scatterers. Assuming $\theta$ is randomly distributed, averaging over the phase difference and applying Landauer's
formula, we arrive at the conductance formula (in
dimensionless quantum units) for 2 scatterers in series, \be
(1+\frac{1}{g})=(1+\frac{1}{g_1})(1+\frac{1}{g_2}),\ee 
where $g_1$ and $g_2$ are typical conductances. In general, for $n$ scatterers in series:
 \be (1+\frac{1}{g})=\prod_{i=1}^{n} (1+\frac{1}{g_i}).
\label{eq:qseries}\ee

As for the formula for adding quantum resistors in parallel,
Arnovas and collaborators~\cite{arnovas} have demonstrated, via a redefinition of a transfer matrix to quantify vertical propagation (as opposed to horizontal propagation in the series case), that the transmission probability is
interchanged with the reflection probability. Therefore,  
to \be (1+g)=\prod_{i=1}^{n} (1+{g_i})
\label{eq:qparallel}\ee for $n$ quantum resistors in parallel, again, assuming the phase randomizes between scatterers. 

In applying these formulas to the Bethe lattice, we assume one scatterer per occupied bond. Using Eqns.~\ref{eq:qseries} and ~\ref{eq:qparallel}, we define the quantum analogs of $S_n$ and $T_n$, or $S_n^q$ and $T_n^q$ as
\be
T_n^q=\prod_i(1+g_i)-1,
\ee
and
\be
S_n^q=\frac{g_0 T_n}{T_n+g_0+1},
\ee
where $g_0$ denotes the typical conductance of an individual scatterer. Note that while there is disorder in terms of the dilution $p$, there is also randomness in the individual conductances. The latter disorder is of the Anderson type, while the former is of the quantum bond percolation type.

\subsection{Bounds on $p_q$ for $k=2,z=3$}
We begin by assuming that the onset of quantum conduction is driven by the
geometric percolation transition. In other words, $p_q=p_c$ and
\be P^q(g)=R{\delta}(g)+(1-R)H^q(g)
\label{assumption} \ee
with
\be
H^q(g)=2pR\int dg_1H^q(g_1){\delta}(g-S^q_1)+p(1-R)\int dg_1dg_2H^q(g_1)H^q(g_2){\delta}(g-S^q_2).\label{eq:hofsigmaquantum}\ee
If an occupied bond is connected to
infinity geometrically via occupied bonds, it is also connected to infinity quantum
mechanically via occupied bonds.  Assuming the quantum conduction transition is
continuous just as the geometric transition, we propose
\be
H^q(g)=\frac{1}{\epsilon^{\gamma}}\bar{H}^q(\frac{g}{\epsilon^\gamma}).
\ee Inserting this ansatz into Eq.~\ref{eq:hofsigmaquantum}
leads to the $\epsilon$ independent terms:
$\mathcal{\bar{H}}^q(s)$ and $\int dx_1 \bar{H}^q(x_1)
\exp(-s(x_1\sigma_0)/(1+\sigma_0))$.  Since these expressions
cannot be equated, the initial assumption of the quantum
conduction transition being driven by the geometry is incorrect,
as expected.  Loosely speaking, quantum interference prevents extended states in narrow channels. In other words, there are not enough occupied bonds participating in the spanning $k=2$ cluster at the transition to warrant a quantum transition. Certainly, the obvious lower bound on $p_q$ is
$p_c$.  To compute a better lower bound on $p_q$, one needs to calculate the quantum
mechanical version of $R$ with the two types of disorder.

We now analyze the fully occupied case to determine if $p_q<1$, i.e. if there exists a quantum conduction transition. For the fully occupied Bethe lattice that there is a
critical value of $g_0$ below which there is no quantum
conduction even for $p=1$. This critical value $g_{0c}$ is given by \be
S_2^q=\frac{g_{0c}[(1+S_2^q)^2-1]}{1+g_{0c}+[(1+S_2^q)^2-1]}.
\ee For $S_2^q<<1$ near the transition (assuming it is continuous), $g_{0c}=1/2$.  For
general $z$, $g_{0c}=1/(z-2)$. This result agrees with
Shapiro~\cite{shapiro}.

We perturb about $p=1$. For $z=3$ and $p=1$, $g_{0c}=1$, we, therefore, choose $g_0=2$ as an example and eventually invoke the expansion parameter $c=1-p<<1$. The conductivity
$g_b$ of a perfect branch is given by: \be
1+\frac{1}{g_b}=(1+\frac{1}{2})(1+\frac{1}{(1+g_b)^{2}-1}),
\ee
yielding $g_b=1$. Following Stinchcombe~\cite{stinchcombe}, we
denote $g^{(n)}$ and $g_{b}^{(n)}$ as the
conductances of the branching network and of any
one of the $z$ branches incident on the origin where
one bond has been removed from the $nth$ shell, respectively. Note that removing one bond from the $nth$ shell does not initiate the removal of other occupied bonds for $k=2,z=3$. For $n\ge2$ \be
1+{g}^{(n)}=(1+{g}_b)^{2}(1+g_b^{(n)}) \ee and
\be
1+\frac{1}{{g}_b^{(n)}}=(1+\frac{1}{2})(1+\frac{1}{(1+{g}_b)(1+{g}_b^{(n-1)})-1}).\ee

These two equations result in
\be
\frac{4}{{g}^{(n)}-3}=\frac{1}{2}+\frac{3}{{g}^{(n-1)}-1}.
\ee
Starting with an initial value of $g^{(1)}=3$, this sequence converges to a finite value of approximately $7$. The number of ways of removing a bond from the $n$th shell
$(n\ge1)$ is $z(z-1)^{n-1}$. Therefore, the average conductivity with a
small concentration $c$ of absent conductors is given by
 \be {g}(c)={g}(c=0)[1-c
\sum_{n=1}^{\infty}z(z-1)^{n-1}(\frac{{g}(c=0)-{g}^{(n)}}{{g}(c=0)})].
\ee
The sum in the above equation diverges as $n\rightarrow\infty$. However, one
can always choose a small enough number $c$ such that $c
\sum_{n=1}^{\infty}z(z-1)^{n-1}(\frac{{g}(c=0)-{g}^{(n)}}{{g}(c=0)})$ remains small. Consequently, there is quantum conduction just below $p=1$, provided $g_0$ is large enough. Therefore, $p_c<p_q<1$ for $g_0>g_{0c}$.

\subsection{$k=3,z=4$ case}
\begin{figure}[bth]
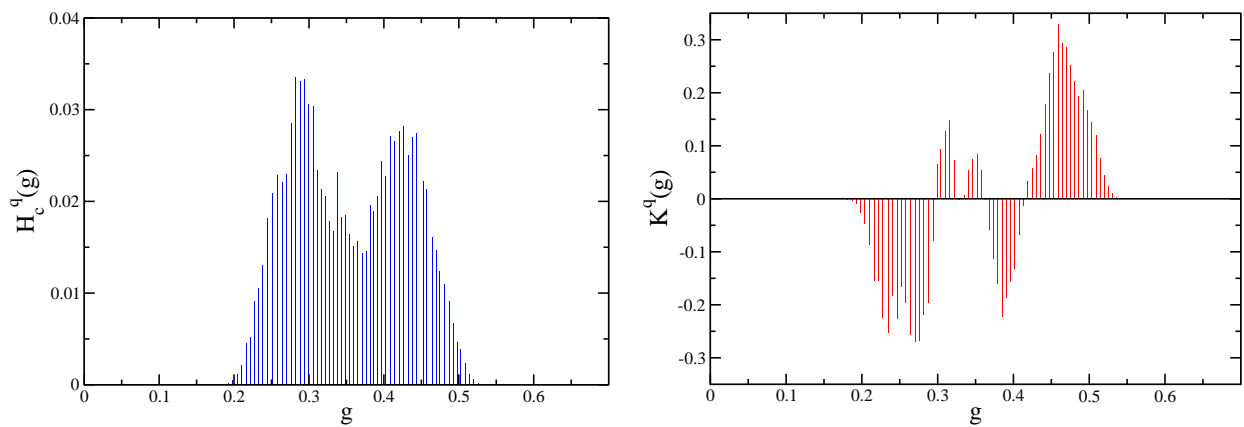

\begin{center}
\includegraphics[width=8cm]{k3.z4.quantum.bond.hofsigma.eps}
\hspace{0.2cm}
\includegraphics[width=8cm]{k3.z4.quantum.bond.kofsigma.eps}
\caption{Left: Plot of $H_c^q(g)$ with binsize $0.005$ for $k=3,z=4$ quantum bond percolation.  Right: Plot of $K^q(g)$ for the same conditions.}
\label{fig:k3z4quantumbond}
\end{center}
\end{figure}

We, again, assume that the onset of quantum conduction is driven by the geometric transition.  This assumption is more plausible for this particular value of $k$ since the onset of the infinite cluster is discontinuous such that one may expect a quantum transition.  More specifically, we assume (1) $p_c=p_q$ and (2)
\be
P^q(g)=R\delta(g)+(1-R)H^q(g),
\label{eq:QPsigma}\ee
where $H^q(g)$ is dictated by the quantum version of Eq.~\ref{eq:hofsigmak3z4}. Assuming $H^q(g)$ has the same scaling form as in the classical $k=3,z=4$ case, we find $\lambda_q=1/2$.  We can also construct the lower bound and upper bounds for the domain of $H_c^q(g)$ using the same arguments in Sec. II and find an upper bound of $(\sqrt{5}-1)/2$ and zero as the lower bound. In Figure~\ref{fig:k3z4quantumbond}, using the same algorithm as in the classical case to solve the nonlinear integral equations, we plot $H_c^q(g)$ and $K^q(g)$.  We find that the average microscopic conductance is
\be
<g>=d_1+d_2\epsilon^{1/2}
\ee
with $d_1=1.379$ and $d_2=4.959$. For the Bethe lattice, the microscopic quantum conductivity, $<\sigma^q>$, is proportional to microscopic quantum conductance and so $<\sigma^q>$ has the same scaling with $\epsilon$. 

The significance of this result is two-fold: (1) we have found a quantum conduction transition with $p_c=p_q$ and (2) we have found a discontinuous onset of quantum conduction in an Anderson model with two types of disorder, one correlated---$k$-core correlated---and one not. Therefore, we have discovered a new universality class for an Anderson-type transition with the classical random first-order phase transition in the geometry driving the quantum transition.
\vspace{1cm}

\section{Discussion}

In the world of metal-insulator transitions, there are typically two effects to consider:  that of electron-electron interactions
and that of disorder.  Typically, disorder-driven (Anderson) transitions~\cite{mirlin}, in the absence of interactions, are
continuous, while interaction-driven transitions in pure systems, such as the Mott-Hubbard transition~\cite{mott}, are discontinuous. Our finding blurs this conventional wisdom in that we have discovered a discontinuous onset in quantum conduction as a function of the $k$-core correlated disorder in the absence of electron-electron interactions. Presumably, there are other geometrical correlations to be constructed and studied---ones that will affect the usually continuous nature of the Anderson transition.

Can this $k$-core disorder be realized in an actual experiment? One motivation for $k$-core (bootstrap) percolation is to capture some aspect of the principle of local mechanical stability in a static, amorphous packing of jammed spheres~\cite{slc}.  Perhaps a quantum analog of this can be realized in low-temperature packings of metallic nanoparticles? An experiment has
already been conducted with
a collection of silver quantum dots sitting atop of a Langmuir monolayer at
room temperature~\cite{heath}.  As the
interparticle spacing decreases by compressing the floating particles together, the electronic
transport goes from hopping to tunnelling to ordinary
metallic transport.
The authors claimed that disorder in the particle size and in the
charging energy
probably does not drive the transition and, instead, argue for a possible
first-order Mott transition at {\em room temperature}. However, in light of the analysis of the onset of classical conduction for $k=3$, we
argue for a possible {\em classical} correlated percolation transition
in conduction. 

Finally, our system contains two types of disorder--the $k$-core dilution disorder and the disorder in the individual conductances. It would be interesting to retain only the $k$-core dilution disorder to realize a $k$-core version of quantum bond percolation. Perhaps then we would also find $p_c=p_q$? We note that Luck and Ashavi~\cite{luck} have analytically investigated quantum conduction on the fully occupied Bethe lattice in the absence of site disorder. They found a band structure in the fully occupied case that shrinks to zero as $z$ is increased.  They did not analyze the dilute case, however. Harris has analyzed quantum uncorrelated bond percolation as the localization transition is approached from below by solving the Schrodinger equation for $E=0$ eigenstates on finite clusters~\cite{harris1,harris2}.  Since no finite clusters exist for $k\ge 3$ on the Bethe lattice, a nontrivial extension of this analysis is required. Such a task, however, should be pursued to discover other models with correlated disorder exhibiting novel metal-insulator transitions. \\

JMS would like to acknowledge funding from NSF-DMR-0645373 and the hospitality of the Aspen Center of Physics where some of this work was completed. The authors would like to thank Ron Maimon for a critical reading of an earlier version of the manuscript.\\

\end{document}